\documentclass{ws-procs975x65}
\usepackage{color}
\usepackage{amsmath}
\usepackage{amssymb}
\usepackage{graphics}

\begin{document}

\title{COSMOLOGICAL BACKREACTION AND THE FUTURE EVOLUTION OF AN ACCELERATING UNIVERSE}

\author{NILOK BOSE$^*$ and A. S. MAJUMDAR}

\address{S. N. Bose National Centre for Basic Sciences,\\
 Block JD, Sector III, Salt Lake, Kolkata 700098, India\\
 $^*$Email: nilok@bose.res.in}
 
\begin{abstract}
We investigate the effect of backreaction due to inhomogeneities on
the evolution of the present universe within the Buchert framework.
Our analysis shows how backreaction from inhomogeneities in the presence
of the cosmic event horizon causes the current acceleration of the
Universe to slow down in the future and even lead in certain cases
to the emergence of a future decelerating epoch.
\end{abstract}

\keywords{Dark Energy; Cosmic Backreaction; Large Scale Structure.}

\bodymatter

\section{Introduction}

The present acceleration of the Universe is well established observationally\cite{perlmutter},
but far from understood theoretically, although there is no dearth
of innovative ideas\cite{sahni}. In recent times there is an upsurge
of interest on studying the effects of inhomogeneities on the expansion
of the Universe and several approaches have been developed to facilitate
this\cite{zala,buchert1,kolb1} and it has been argued\cite{rasanen1}
that backreaction from inhomogeneities from the era of structure formation
could lead to an accelerated expansion of the Universe.

\section{The Backreaction Framework}

In the framework developed by Buchert\cite{buchert1,weigand} for
a compact spatial domain $\mathcal{D}$ the scale-factor, $a_{\mathcal{D}}(t)=\left(\frac{|\mathcal{D}|_{g}}{|\mathcal{D}_{i}|_{g}}\right)^{1/3}$,
encodes the average stretch of all directions of the domain, where
$|\mathcal{D}|_{g}$ is the volume of $\mathcal{D}$. Using the Einstein
equations, with a pressure-less fluid source, we get the following
equations\cite{buchert1,weigand} 
\begin{eqnarray}
3\frac{\ddot{a}_{\mathcal{D}}}{a_{\mathcal{D}}} & = & -4\pi G\left\langle \rho\right\rangle _{\mathcal{D}}+\mathcal{Q}_{\mathcal{D}}+\Lambda\label{eq:1a}\\
3H_{\mathcal{D}}^{2} & = & 8\pi G\left\langle \rho\right\rangle _{\mathcal{D}}-\frac{1}{2}\mathcal{\left\langle R\right\rangle }_{\mathcal{D}}-\frac{1}{2}\mathcal{Q}_{\mathcal{D}}+\Lambda\label{eq:1b}
\end{eqnarray}
Here the average of the scalar quantities on the domain $\mathcal{D}$
is defined as, $\left\langle f\right\rangle {}_{\mathcal{D}}(t)=|\mathcal{D}|_{g}^{-1}\int_{\mathcal{D}}fd\mu_{g}$
and where $\rho$, $\mathcal{R}$ and $H_{\mathcal{D}}$ denote the
local matter density, the Ricci-scalar, and the domain dependent Hubble
rate $H_{\mathcal{D}}=\dot{a}_{\mathcal{D}}/a_{\mathcal{D}}$ respectively.
The kinematical backreaction $\mathcal{Q_{D}}=\frac{2}{3}\left(\left\langle \theta^{2}\right\rangle _{\mathcal{D}}-\left\langle \theta\right\rangle _{\mathcal{D}}^{2}\right)-2\sigma_{\mathcal{D}}^{2}$
where $\theta$ is the local expansion rate and $\sigma^{2}=1/2\sigma_{ij}\sigma^{ij}$
is the squared rate of shear.

The ``global'' domain $\mathcal{D}$ is assumed to be separated
into subregions and following Ref.~\refcite{weigand} we work with
only two subregions. Clubbing those parts of $\mathcal{D}$ which
consist of initial overdensity as $\mathcal{M}$ (called `wall'),
and those with initial underdensity as $\mathcal{E}$ (called `void'),
such that $\mathcal{D}=\mathcal{M}\cup\mathcal{E}$, one obtains 
\begin{equation}
\frac{\ddot{a}_{\mathcal{D}}}{a_{\mathcal{D}}}=\lambda_{\mathcal{M}}\frac{\ddot{a}_{\mathcal{M}}}{a_{\mathcal{M}}}+\lambda_{\mathcal{E}}\frac{\ddot{a}_{\mathcal{E}}}{a_{\mathcal{E}}}+2\lambda_{\mathcal{M}}\lambda_{\mathcal{E}}(H_{\mathcal{M}}-H_{\mathcal{E}})^{2}\label{eq:6}
\end{equation}
Here $\sum_{\ell}\lambda_{\ell}=\lambda_{\mathcal{M}}+\lambda_{\mathcal{E}}=1$,
with $\lambda_{\mathcal{M}}=|\mathcal{M}|/|\mathcal{D}|$ and $\lambda_{\mathcal{E}}=|\mathcal{E}|/|\mathcal{D}|$,
and $a_{\mathcal{M}}$ , $H_{\mathcal{M}}$ and $a_{\mathcal{E}}$,
$H_{\mathcal{E}}$ are the scale factors and Hubble parameters of
the $\mathcal{M}$ and $\mathcal{E}$ subdomains respectively.

\section{Effect of event horizon}

We consider the universe once the present stage of acceleration sets
in and try to see the effect of backreaction in the presence of the
cosmic event horizon (first presented in Ref.~\refcite{bose}). We
can write the equation of the event horizon $r_{h}$, to a good approximation
by 
\begin{equation}
r_{h}=a_{\mathcal{D}}\int_{t}^{\infty}\frac{dt'}{a_{\mathcal{D}}(t')}\label{eq:7}
\end{equation}
The void-wall symmetry of Eq.(\ref{eq:6}) ensures that the conclusions
are similar whether one chooses to define the event horizon with respect
to the wall or with respect to the void.

We assume that the scale-factors of the regions $\mathcal{E}$ and
$\mathcal{M}$ are, respectively, given by $a_{\mathcal{E}}=c_{\mathcal{E}}t^{\alpha}$
and $a_{\mathcal{M}}=c_{\mathcal{M}}t^{\beta}$ where $\alpha$, $\beta$,
$c_{\mathcal{E}}$ and $c_{\mathcal{M}}$ are constants. Since an
event horizon forms, only those regions of $\mathcal{D}$ that are
within the event horizon are causally accessible to us. Therefore
we have to introduce an apparent volume fraction of $\mathcal{M}$
which is defined as $\lambda_{\mathcal{M}_{h}}=\frac{|\mathcal{M}|_{g}}{\frac{4}{3}\pi r_{h}^{3}}=\frac{c_{\mathcal{M}_{h}}^{3}t^{3\beta}}{r_{h}^{3}}$,
where $c_{\mathcal{M}_{h}}^{3}=3c_{\mathcal{M}}^{3}|\mathcal{M}_{i}|_{g}/4\pi$
is a constant. Normalizing the total accessible volume in the presence
of the horizon we can write $\lambda_{\mathcal{E}_{h}}=1-\lambda_{\mathcal{M}_{h}}$,
where $\lambda_{\mathcal{E}_{h}}$ is the apparent volume fraction
of the sub-domain $\mathcal{E}$. It hence follows that global acceleration
equation \eqref{eq:6} is now given by 

\begin{eqnarray}
\frac{\ddot{a}_{\mathcal{D}}}{a_{\mathcal{D}}} & = & \frac{c_{\mathcal{M}_{h}}^{3}t^{3\beta}}{r_{h}^{3}}\frac{\beta(\beta-1)}{t^{2}}+\left(1-\frac{c_{\mathcal{M}_{h}}^{3}t^{3\beta}}{r_{h}^{3}}\right)\frac{\alpha(\alpha-1)}{t^{2}}\nonumber \\
 &  & +2\frac{c_{\mathcal{M}_{h}}^{3}t^{3\beta}}{r_{h}^{3}}\left(1-\frac{c_{\mathcal{M}_{h}}^{3}t^{3\beta}}{r_{h}^{3}}\right)\left(\frac{\beta}{t}-\frac{\alpha}{t}\right)^{2}\label{eq:11-1}
\end{eqnarray}

\begin{figure}
\begin{center}
\psfig{file=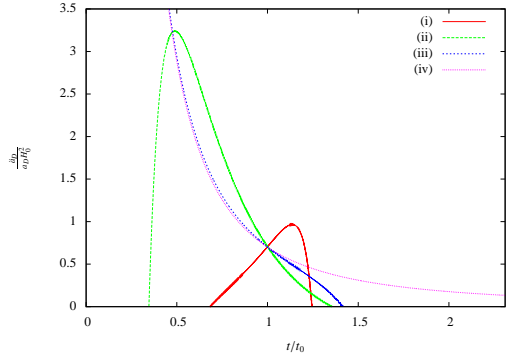,width=6.6cm}
\end{center}
\caption{The dimensionless global acceleration parameter $\frac{\ddot{a}_{\mathcal{D}}}{a_{\mathcal{D}}H_{0}^{2}}$,
plotted vs. time (in units of $t/t_{0}$ with $t_{0}=4.3\times10^{17}\mathrm{s}$
). The parameter values used are: (i) $\alpha=0.995$, $\beta=0.5$,
(ii) $\alpha=0.984$, $\beta=0.5,$ (iii) $\alpha=1.02$, $\beta=0.66$,
(iv) $\alpha=1.02$, $\beta=0.66$.}
\label{fig1}
\end{figure}

The current acceleration of the Universe ensures the formation of
the event horizon, so $r_{h}$ defined by \eqref{eq:7} will be finite
valued, thus enabling us to rewrite \eqref{eq:7} as

\begin{equation}
\dot{r_{h}}=\frac{\dot{a}_{\mathcal{D}}}{a_{\mathcal{D}}}r_{h}-1\label{eq:13}
\end{equation}

We can therefore solve numerically the set of coupled differential
equations \eqref{eq:13} and \eqref{eq:11-1} by using as an `initial
condition' the observational constraint $q_{0}=-0.7$, where $q_{0}$
is the current value of the deceleration parameter.

\section{Discussions and Conclusions}

In Fig. 1 the curves (i) and (iii) are for the case when an event
horizon is included, and curves (ii) and (iv) correspond to the case
without an event horizon. We see that whether $\alpha<1$ or $\alpha>1$,
the acceleration always becomes negative in the future when we include
the event horizon (curves (i) and (iii)), whereas the acceleration
only becomes negative for $\alpha<1$ when we don't include the horizon
in our calculations (curve (ii)). We also see that the deceleration
is much faster when we include the event horizon, the reason for that
could be that the inclusion of the event horizon somehow decreases
the available volume of the underdense region $\mathcal{E}$ which
causes the overdense region $\mathcal{M}$ to start dominating much
earlier and leads to global deceleration much more quickly

Our results indicate the fascinating possibility of backreaction being
responsible for the slowing down of the current accleration and in
some cases cause a transition to a future declerated era, no matter
what the cause of the current acceleration.

\bibliographystyle{ws-procs975x65}

\end{document}